\documentclass{Interspeech}

\interspeechcameraready

\title{Evaluating Logit-Based GOP Scores for Mispronunciation Detection}

\author{Aditya Kamlesh}{Parikh}
\author{Cristian}{Tejedor-Garcia}
\author{Catia}{Cucchiarini}
\author{Helmer}{Strik}

\affiliation[nocounter]{Centre for Language Studies}{Radboud University}{the Netherlands}
\email{aditya.parikh@ru.nl, cristian.tejedorgarcia@ru.nl, catia.cucchiarini@ru.nl, helmer.strik@ru.nl}

\usepackage{comment}
\usepackage{graphicx}

\usepackage{tabularx}
\usepackage{booktabs}
\usepackage{multirow}
\usepackage{subfig}
\usepackage{comment}
\usepackage[utf8]{inputenc} 
\usepackage{graphicx} 
\usepackage{amsmath} 
\usepackage{amssymb} 
\usepackage{array} 
\usepackage{enumitem} 
\usepackage{tipa} 
\usepackage{booktabs}

\begin{document}

\maketitle

\begin{abstract}
Pronunciation assessment relies on goodness of pronunciation (GOP) scores, traditionally derived from softmax-based posterior probabilities. However, posterior probabilities may suffer from overconfidence and poor phoneme separation, limiting their effectiveness. This study compares logit-based GOP scores with probability-based GOP scores for mispronunciation detection. We conducted our experiment on two L2 English speech datasets spoken by Dutch and Mandarin speakers, assessing classification performance and correlation with human ratings. Logit-based methods outperform probability-based GOP in classification, but their effectiveness depends on dataset characteristics. The maximum logit GOP shows the strongest alignment with human perception, while a combination of different GOP scores balances probability and logit features. The findings suggest that hybrid GOP methods incorporating uncertainty modeling and phoneme-specific weighting improve pronunciation assessment.
\end{abstract}

\section{Introduction}
\label{sec:Introduction}

In today's interconnected world, globalization has led to increased movement across borders for work, education, and other opportunities. For individuals who are adapting to a new linguistic environment, learning the local language is essential for social integration, career advancement, and overall well-being \cite{kuschel2023combining}. Effective communication goes beyond knowing the vocabulary and grammar: A clear pronunciation is equally important, as it impacts intelligibility, confidence, and the ability to engage in meaningful conversations \cite{reading101017,jenkins2000phonology}. Poor pronunciation can lead to misunderstandings, impede effective social interactions and even create barriers in academic and professional settings \cite{zossadult}. However, mastering pronunciation in a second language (L2) can be challenging. Differences between the first language (L1) and the target language often result in persistent pronunciation errors. These difficulties are further compounded by the limited tools and resources available to language instructors for providing personalized pronunciation feedback.

To address these challenges, Computer-Assisted Pronunciation Training (CAPT) systems have gained popularity \cite{hsu2024examination,Amrate2024ComputerassistedPT}. A key component of these systems is Mispronunciation Detection and Diagnosis (MDD), which helps learners identify and correct pronunciation errors in real time \cite{alrashoudi2025improving}. Phoneme-level assessment, in particular, provides more precise feedback than broader word- or sentence-level evaluations, allowing learners to focus on specific areas for improvement \cite{cao2024framework}. One of the most widely used methods for detecting phoneme-level mispronunciations is the goodness of pronunciation (GOP) score \cite{witt2000use}.

GOP was initially introduced as a measure of pronunciation quality, estimating the probability of a phoneme and comparing it against a predefined threshold to flag mispronunciations \cite{kanters2009goodness}. Over time, several enhancements have improved its accuracy. Weighted-GOP \cite{doremalen10_interspeech} adjusts phoneme scores based on linguistic and acoustic factors, prioritizing phonemes prone to mispronunciation. Lattice-based GOP \cite{5451809} considers multiple pronunciation possibilities using phoneme lattices, yielding more robust confidence scores. Context-aware GOP \cite{shi2020context} incorporates phoneme transitions and durations to better capture natural pronunciation variations. More recently, multidimensional GOP features \cite{do2024acoustic,cao2024framework,gong_gopt} have been introduced, leveraging richer feature representations beyond simple probability thresholds for more precise mispronunciation detection. 

Despite these advancements, GOP methods face challenges in data availability and computational efficiency \cite{elkheir2023mispronunciation}. Annotated pronunciation data, including prosody and fluency scores, requires expert evaluation, making large-scale collection expensive. Additionally, traditional GOP methods scale poorly with phoneme size because of an increasing number of features and increasing computational costs. This highlights the need for a fast, robust, and non-trainable GOP computation method.

More recently, foundation models \cite{babu2021xls,10.1109/TASLP.2021.3122291} trained on massive amounts of data have been used to improve MDD. These models can be fine-tuned with significantly less data, addressing some of the data availability constraints faced by traditional GOP-based methods. GOP scores can be derived from a forced alignment using Connectionist Temporal Classification (CTC)-based phoneme recognition models, which rely on posterior probabilities generated by acoustic models. These probabilities are obtained through softmax normalization of model logits, which is a widely adopted method \cite{sudhakara2019improved}. However, softmax-based probability estimates suffer from overconfidence \cite{pearce2021understanding,wei2022mitigating}, inflating confidence in incorrect phoneme predictions and reducing the granularity needed to detect subtle mispronunciations. This issue is particularly problematic in phoneme recognition for children's speech and non-native speakers, where articulatory deviations are very common \cite{xie2020comparing,preston2015perception}.

To address these limitations, we propose a logit-based GOP method that directly utilizes raw logits from CTC-based models rather than softmax-normalized probabilities. Logits retain more discriminative information and avoid the gradient saturation problem inherent in softmax-based scoring \cite{li2020learning}. We explore four logit-based metrics to enhance mispronunciation detection: Maximum Logit (GOP\textsubscript{MaxLogit}), Mean Logit Margin \cite{10147340} (GOP\textsubscript{Margin}), Logit Variance (GOP\textsubscript{LogitVariance}), and combined (hybrid) logit-probability GOP\textsubscript{Combined} scores. This novel approach provides a fast, robust, and non-trainable solution, crucial for real-time phoneme assessment.

Our method builds upon prior work on uncertainty quantification in pronunciation assessment, such as \cite{yeo2023speech}, which applies GOP-based scores to dysarthric speech. However, prior studies have not explicitly investigated the use of raw logits in GOP calculations. Our approach fills this gap by utilizing logit-based metrics to improve accuracy and reliability in pronunciation assessment.
Our leading research question (RQ) is: To what extent does a logit-based GOP score enhance mispronunciation detection and improve correlation with human rater scores compared to traditional softmax-based GOP scores?

\section{Methodology}
\subsection{Definition of GOP}

The GOP score, first introduced by Witt and Young~\cite{witt2000use}, quantifies pronunciation quality by comparing the likelihood of a hypothesized phoneme to competing alternatives. For a phoneme \( p \) aligned to an audio segment, the original GOP formulation computes:
\begin{equation}
    \text{GOP}_{\text{original}}(p) = \log \frac{P(\mathbf{X} | p)}{ \frac{1}{N} \sum_{q \in \mathcal{Q}} P(\mathbf{X} | q) }
    \label{eq:gop_original}
\end{equation}
where \( P(\mathbf{X} | p) \) is the likelihood of the acoustic features \( \mathbf{X} \) given phoneme \( p \), and \( \mathcal{Q} \) represents competing phonemes.

With deep neural networks (DNNs), GOP is derived from posterior probabilities using the negative log of the mean softmax output over aligned frames \cite{hu2013new}:
\begin{equation}
    \text{GOP}_{\text{DNN}}(p) = -\log \left( \frac{1}{T} \sum_{t=1}^T P(p | \mathbf{x}_t) \right)
    \label{eq:gop_dnn}
\end{equation}
where \( P(p | \mathbf{x}_t) \) is the softmax probability of phoneme \( p \) at frame \( t \) and $T$ is the total number of frames in the phoneme segment.  This equation interprets the mean softmax probability of the target phoneme as a probabilistic measure of pronunciation quality. This approach inherits softmax limitations such as overconfidence and gradient saturation.

\subsection{Logit-Based GOPS}
To address softmax limitations, we propose four novel metrics using raw logits, defined below.

\subsubsection{GOP\textsubscript{MaxLogit}}
This metric captures the model's peak confidence in the target phoneme \( p \) across aligned frames \( t_1 \) to \( t_2 \):
\begin{equation}
    \text{GOP}_{\text{MaxLogit}}(p) = \max_{t \in [t_1, t_2]} \mathbf{l}_t^{(p)}
    \label{eq:gop_maxlogit}
\end{equation}
where \( \mathbf{l}_t^{(p)} \) is the logit for phoneme \( p \) at frame \( t \). It identifies unambiguous articulations but may emphasize transient spikes.

\subsubsection{GOP\textsubscript{Margin}}
This measure quantifies the average superiority of the target phoneme over its strongest competitor. For each frame, we compute the difference (or margin) between the target logit and the highest competing logit. The average of these margins over the segment indicates how well-separated the target phoneme is from other phonemes. This helps in cases where pronunciation errors cause phoneme confusion, which may not always be reflected in probability scores.

\begin{equation}
    \text{GOP}_{\text{Margin}}(p) = \frac{1}{T} \sum_{t=t_1}^{t_2} \left( \mathbf{l}_t^{(p)} - \max_{k \neq p} \mathbf{l}_t^{(k)} \right),
    \label{eq:gop_margin}
\end{equation}

\subsubsection{\texorpdfstring{GOP\textsubscript{VarLogit}}{GOP VarLogit}}

This metric measures the \textit{variability} of the model's confidence in predicting the target phoneme across timeframes. 

\vspace{-0.2cm}
\begin{equation}
    \text{GOP}_{\text{VarLogit}}(p) = \frac{1}{T} \sum_{t=t_1}^{t_2} \left( \mathbf{l}_t^{(p)} - \mu_p \right)^2, \quad \mu_p = \frac{1}{T} \sum_{t=t_1}^{t_2} \mathbf{l}_t^{(p)},
    \label{eq:gop_variance}
\end{equation}

It is computed as the variance of the raw logit values associated with the target phoneme. A low logit variance suggests that the model consistently assigns similar confidence levels to the phoneme across frames, indicating a stable and confident recognition. Conversely, a high logit variance implies fluctuating confidence, which may occur due to acoustic distortions, coarticulation effects, or phonetic ambiguity.

\subsubsection{\texorpdfstring{GOP\textsubscript{Combined}}{GOP Combined}} 
This hybrid metric is designed to use the strengths of both logit-based and probability-based approaches to pronunciation assessment. It integrates the Mean Logit Margin, which quantifies the relative confidence of the target phoneme against competing phonemes, and the traditional $\text{GOP}_{\text{DNN}}$.  

\vspace{-0.2cm}
\begin{equation}
    \text{GOP}_{\text{Combined}}(p) = \alpha \cdot \text{GOP}_{\text{Margin}}(p) - (1-\alpha) \cdot \text{GOP}_{\text{DNN}}(p)
    \label{eq:gop_combined}
\end{equation}
where \( \alpha \in [0,1] \) balances contributions. By combining these two metrics, the combined score may provide a more balanced assessment of pronunciation quality, mitigating the weaknesses of each individual measure. This hybrid approach ensures that both posterior probability (via $\text{GOP}_\text{DNN}$) and phoneme separability (via $\text{GOP}_\text{margin}$) contribute to the final score, making it more sensitive to pronunciation deviations while reducing the impact of softmax-related limitations.


\subsection{GOP Calculations}

In our study for forced alignment, we utilized the CTC segmentation algorithm \cite{kurzinger2020ctc}, which uses an end-to-end CTC-based phoneme recognition model to determine phoneme boundaries. For the CTC-based acoustic model for phoneme recognition, we utilized an open-source fine-tuned phoneme recognition Wav2vec2.0 model\footnote{\tiny\url{https://huggingface.co/facebook/wav2vec2-xlsr-53-espeak-cv-ft}} based on \cite{xu22b_interspeech}.

\subsection{Datasets}
\label{dataset-info}


To address our RQ, we conducted experiments using two L2 English speech datasets: My Pronunciation Coach (MPC) \cite{cucchiarini2012my} and SpeechOcean762 \cite{zhang2021speechocean762}. MPC comprises recordings of Dutch children speaking English, while the latter includes speech from Mandarin-speaking adults and children. Given its high acoustic variability—resulting from L1 transfer and inconsistent phoneme realizations \cite{gretter21_interspeech}—non-native children's speech was a primary focus, as it presents a particularly challenging testbed for pronunciation assessment.

MPC \cite{cucchiarini2012my} contains speech from 124 Dutch secondary school students. Each recording includes 53 English words and 53 sentences covering a wide range of phonemes. Recordings are categorized into quality groups: Excellent, OK, Doubtful and Overloud. For this study, we used 50 OK and 21 Excellent sessions, totalling 3,130 utterances from 71 speakers (38 males, 33 females).
Since MPC lacks annotated mispronunciations, we introduced simulated pronunciation errors by modifying phoneme sequences. Common substitutions include replacing /\textipa{\dh}/ → /\textipa{d}/, /\textipa{T}/ → /\textipa{s}/, /\textipa{\ae}/ → /\textipa{e}/, and diphthong simplifications such as /\textipa{eI}/ → /\textipa{e:}/.

SpeechOcean762 \cite{zhang2021speechocean762} is an open-source corpus for pronunciation assessment, containing 5,000 English utterances from 250 native Mandarin speakers (125 adults, 125 children). Each utterance is annotated by five experts at the sentence, word, and phoneme levels, with 3,401 phonemes labelled as mispronunciations. We used all 5,000 utterances in our experiments.\footnote{\tiny{\url{https://github.com/Aditya3107/GOP_logit.git}}}

\subsection{Evaluation Metrics}
\label{metricsinfo}


We assessed model performance using accuracy, precision, recall, F1-score, and Matthews Correlation Coefficient (MCC). Given the class imbalance in both datasets, we optimized the GOP threshold by selecting the percentile that maximized MCC. Additionally, we reported the ROC AUC score at this threshold to evaluate classification effectiveness.

In addition, in order to analyze GOP score distributions, we used violin plots to compare posterior probability-based, logit-based, and hybrid GOP scores across correct and mispronounced phonemes. This visualization helps determine whether a given GOP scoring method provides a clear phoneme separation, a key factor in pronunciation assessment.

The Speechocean762 dataset includes human-annotated phoneme accuracy scores. Following prior research \cite{zhang2021speechocean762}, we applied a second-order polynomial regression to model the relationship between GOP and human phoneme accuracy ratings. Performance was evaluated using Pearson Correlation Coefficient (PCC) and Mean Squared Error (MSE) to quantify prediction accuracy on the test set. Finally, phoneme-level mispronunciation error rates of the Speechocean762 dataset were also analyzed using a bar plot, comparing the GOP method with the highest PCC correlation to human-rated phoneme accuracy.

\section{Results}


Table \ref{tab:mpc-table} presents the evaluation scores for posterior probability based (first column), logit-based (second to fourth columns) and hybrid GOP scores (last column) on the MPC dataset. 
Of all these measures, \(\text{GOP}_{\text{Margin}}\) achieves the highest accuracy (0.851), MCC (0.347). It also outperforms other approaches in F1-score (0.415) and precision (0.347), ensuring better mispronunciation detection while maintaining precision. However, \(\text{GOP}_{\text{MaxLogit}}\) achieves the highest AUC at MCC\textsubscript{max} (0.736), making it the most effective in distinguishing correctly pronounced and mispronounced phonemes.  The measure \( \text{GOP}_{\text{DNN}}\) shows the highest recall (0.929) but has the lowest precision (0.184), indicating it detects most mispronunciations but lacks specificity. 
\(\text{GOP}_{\text{MaxLogit}}\) shows moderate performance, achieving an accuracy of 0.590 and an MCC of 0.286.  
Its performance is slightly better than \(\text{GOP}_{\text{DNN}}\) but still lower than \(\text{GOP}_{\text{Margin}}\) in most metrics.  The \(\text{GOP}_{\text{Combined}}\) score achieves an accuracy of 0.82 and MCC of 0.314, showing a good balance.  Its AUC at MCCmax (0.704) is competitive, suggesting it may be a promising hybrid approach.  

\begin{table}[ht!]
\caption{Performance analysis on the MPC dataset}
\vspace{-0.3cm}
\label{tab:mpc-table}
\scriptsize
\setlength{\tabcolsep}{3pt} 
\begin{tabularx}{\columnwidth}{@{}l*{5}{>{\centering\arraybackslash}X}@{}}
\toprule
                       & $\text{GOP}_{\text{DNN}}$ & $\text{GOP}_{\text{MaxLogit}}$ & $\text{GOP}_{\text{Margin}}$ & $\text{GOP}_{\text{VarLogit}}$ & $\text{GOP}_{\text{Combined}}$ \\
\midrule
Accuracy               & 0.572                  & 0.590                  & \textbf{0.851}  & 0.461                   & 0.820              \\
Precision              & 0.184                  & 0.189                  & \textbf{0.347}  & 0.146                   & 0.297              \\
Recall                 & \textbf{0.929}         & 0.919                  & 0.515           & 0.882                   & 0.559              \\
F1                     & 0.307                  & 0.314                  & \textbf{0.415}  & 0.250                   & 0.388              \\
MCC                    & 0.279                  & 0.286                  & \textbf{0.342}  & 0.184                   & 0.314              \\
AUC MCC\textsubscript{max}  & 0.730         & \textbf{0.736}                  & 0.702  & 0.648                   & 0.704              \\
\bottomrule
\end{tabularx}
\end{table}


Table \ref{tab:speechocean-table} shows the evaluation results of different GOP-based pronunciation assessment scores of the SpeechOcean762 dataset. This table also includes PCC and MSE scores since 
the SpeechOcean762 dataset includes human-annotated phoneme accuracy ratings, allowing us to evaluate the correlation between GOP and expert scores.
GOP\textsubscript{DNN} achieves the highest accuracy (0.947), precision (0.333), and MCC (0.367), showing that it effectively separates correctly pronounced and mispronounced phonemes. However, its PCC scores (0.278 for low confidence and 0.295 for high confidence) are significantly lower than other logit-based approaches. This suggests that while GOP\textsubscript{DNN} can classify pronunciation errors well, it does not align well with human-annotated phoneme scores, making it less reliable for subjective pronunciation assessment. GOP\textsubscript{MaxLogit} achieves the highest PCC scores (0.442 for low confidence and 0.456 for high confidence), outperforming all other GOP metrics in correlating with human ratings. This indicates that maximum logit values capture pronunciation quality in a way that aligns better with human perception compared to posterior probability-based GOP. It also achieves a strong AUC at MCC\textsubscript{max} (0.754), reinforcing its reliability as a GOP metric. GOP\textsubscript{Margin} shows the weakest overall performance, with an MCC of only 0.174 and lower PCC scores (0.173 for low confidence and 0.191 for high confidence).

\begin{table}[ht!]
\caption{Performance analysis on the SpeechOcean762 dataset}
\vspace{-0.3cm}
\label{tab:speechocean-table}
\scriptsize
\setlength{\tabcolsep}{3pt} 
\begin{tabularx}{\columnwidth}{@{}l*{5}{>{\centering\arraybackslash}X}@{}}
\toprule
                       & $\text{GOP}_{\text{DNN}}$ & $\text{GOP}_{\text{MaxLogit}}$ & $\text{GOP}_{\text{Margin}}$ & $\text{GOP}_{\text{VarLogit}}$ & $\text{GOP}_{\text{Combined}}$ \\
\midrule
Accuracy               & \textbf{0.947}         & 0.925                  & 0.741           & 0.894                   & 0.843              \\
Precision              & \textbf{0.333}         & 0.257                  & 0.089           & 0.195                   & 0.139              \\
Recall                 & 0.466                  & 0.571                  & \textbf{0.672}  & 0.621                   & 0.642              \\
F1                     & \textbf{0.388}         & 0.354                  & 0.157           & 0.297                   & 0.228              \\
MCC                    & \textbf{0.367}         & 0.350                  & 0.174           & 0.308                   & 0.247              \\
AUC MCC\textsubscript{max}  & 0.715         & \textbf{0.754}         & 0.708           & 0.763                   & 0.747              \\
\hline
PCC (low conf)         & 0.278                  & \textbf{0.442}         & 0.173           & 0.341                   & 0.303              \\
PCC (high conf)        & 0.295                  & \textbf{0.456}         & 0.191           & 0.357                   & 0.319              \\
MSE                    & 0.124                  & \textbf{0.109}         & 0.131           & 0.120                   & 0.123              \\
\bottomrule
\end{tabularx}
\end{table}

While margin-based GOP obtained most of the best metric scores in the MPC dataset, it does not generalize well to SpeechOcean762, possibly due to differences in speaker demographics and phoneme variability. In Table \ref{tab:speechocean-table}  we see that GOP\textsubscript{VarLogit} achieves high recall (0.621) but has weak MCC (0.308) and low precision (0.195), suggesting it works well for detecting mispronunciations but lacks robustness in classification. GOP\textsubscript{Combined} achieves an MCC of 0.247 and PCC scores (0.303 for low confidence and 0.319 for high confidence), indicating a balance between probability and logit-based features.

\begin{figure*}[ht!]
    \centering
    \includegraphics[width=0.8\textwidth]{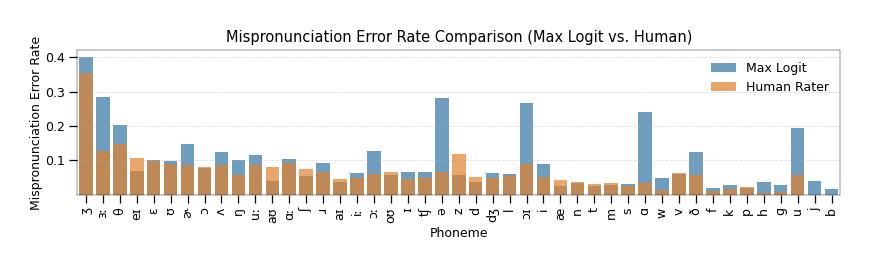}
    \vspace{-0.6cm}
    \caption{Comparison of mispronunciation error rates by phoneme (Max Logit vs. human rater) in the SpeechOcean762 dataset. Blue bars show $\text{GOP}_{\text{MaxLogit}}$-predicted error rates, while red bars indicate human-rated phoneme accuracy.}
    \label{fig:maxlogit_vs_human}
\end{figure*}

\begin{figure}[ht!]
    \centering
    \includegraphics[width=0.35\textwidth]{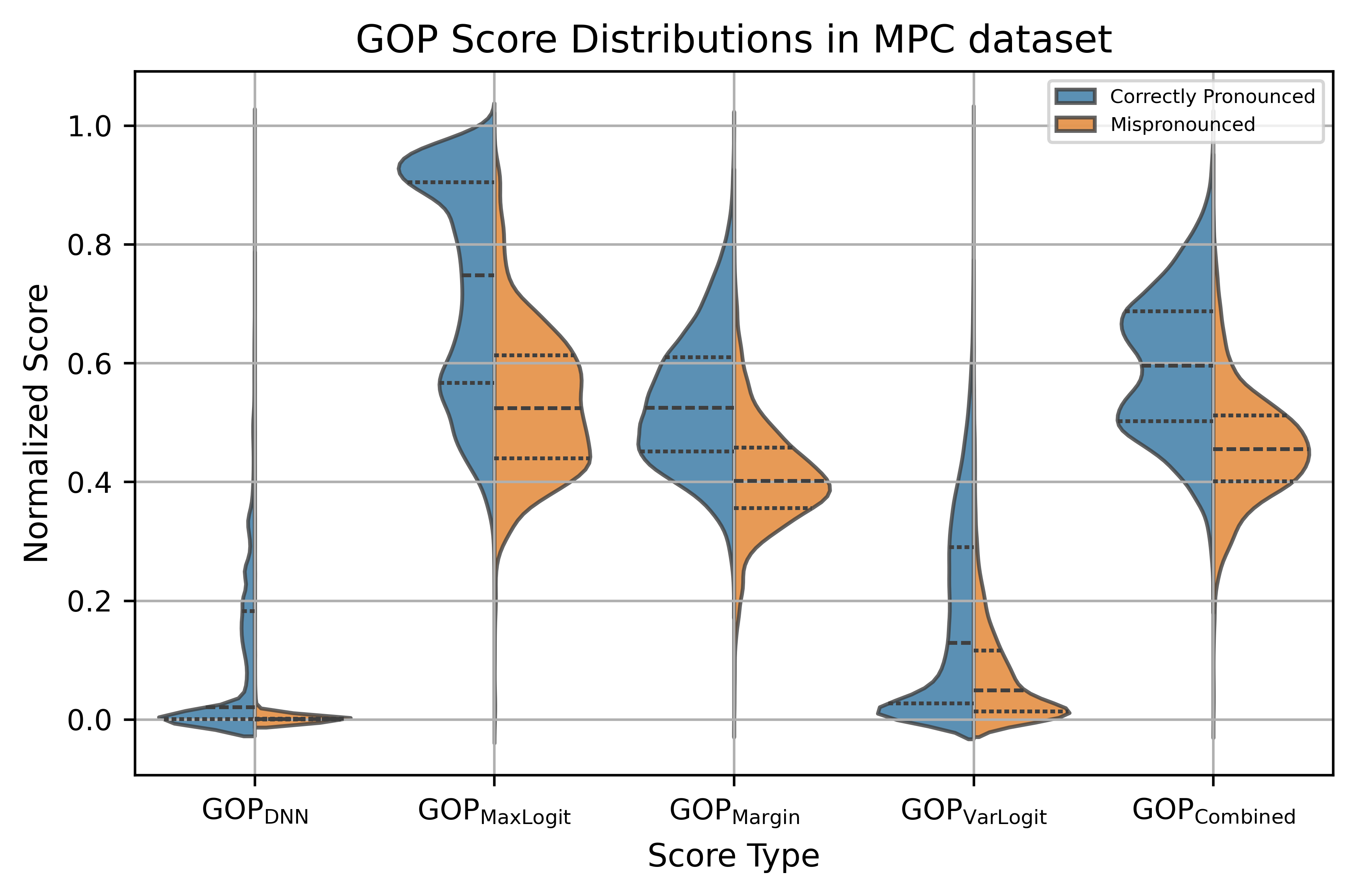} %
    \includegraphics[width=0.35\textwidth]{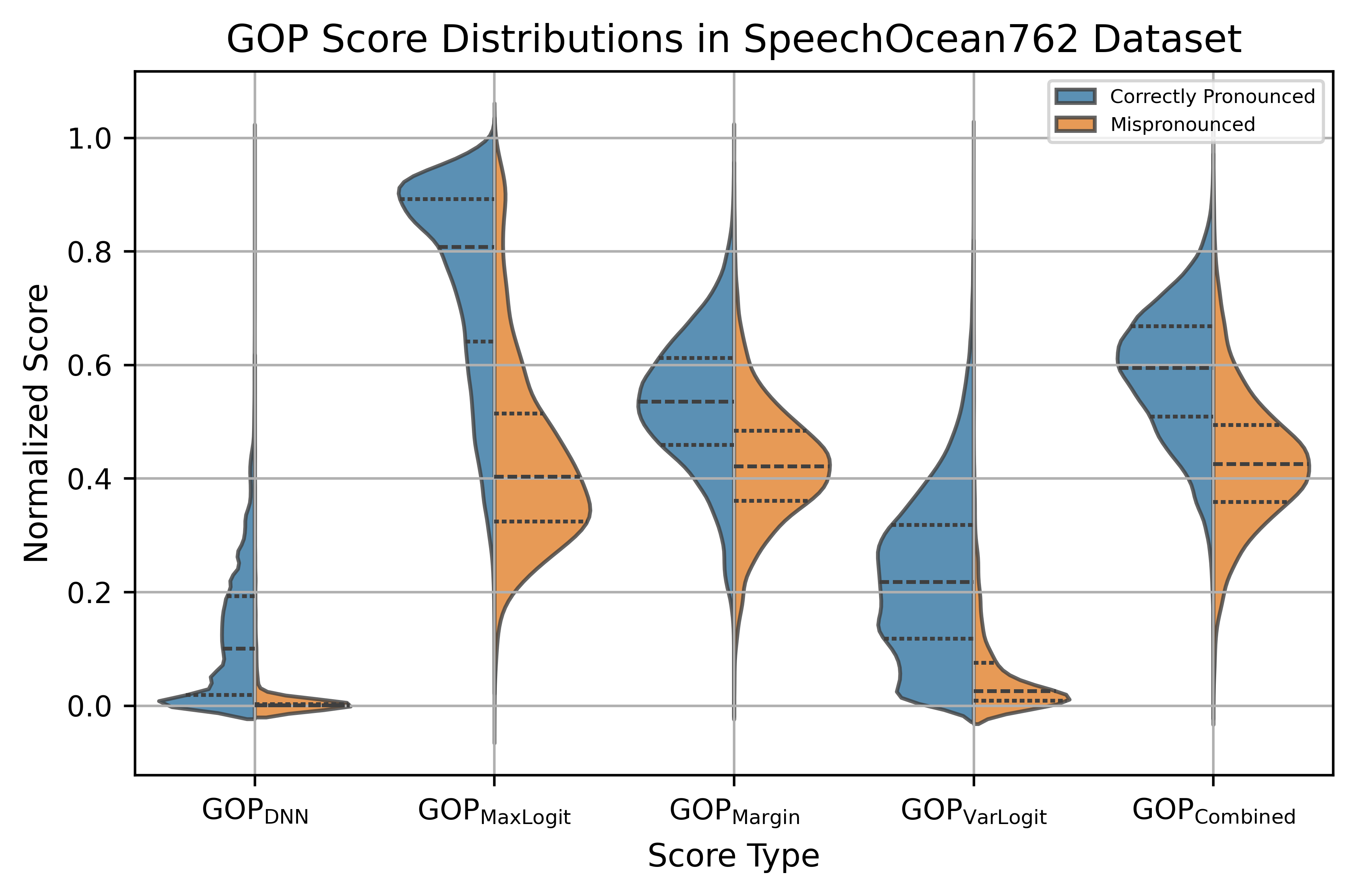}
    \vspace{-0.3cm}
    
        \caption{Comparison of GOP score distributions across MPC and Speechocean762 datasets.}
    \label{fig:gop_violin}
\end{figure}

To better interpret the performance results, Figure \ref{fig:gop_violin} visualizes the distribution of correctly pronounced and mispronounced phonemes across both datasets. GOP\textsubscript{DNN} shows a wide distribution overlap between correct and mispronounced phonemes in both cases (both graphs of Figure \ref{fig:gop_violin}), with close means and high variance, making it less effective for error distinction.
In contrast, GOP\textsubscript{MaxLogit} achieves  better separation, especially in MPC (first graph of Figure \ref{fig:gop_violin}), though overlap increases in SpeechOcean762 (second graph of Figure \ref{fig:gop_violin}), leading to higher variability. 
The best-performing score in MPC, GOP\textsubscript{Margin}, effectively classifies children's speech (first graph of Figure \ref{fig:gop_violin}), but struggles with greater distribution overlap in SpeechOcean762 (second graph of Figure \ref{fig:gop_violin}).
GOP\textsubscript{VarLogit} shows high variability and significant overlap in both cases, making it unreliable. 
GOP\textsubscript{Combined} balances probability- and logit-based approaches across both datasets, reducing overlap compared to GOP\textsubscript{DNN}. It presents a trade-off between classification accuracy from posterior probabilities and human correlation from logit-based scores. While we incorporated  GOP\textsubscript{Margin} in GOP\textsubscript{Combined}, GOP\textsubscript{MaxLogit} could also be considered, leaving the optimal choice undecided.

Finally, to analyze the alignment between GOP-based mispronunciation detection and human-rated phoneme accuracy, we investigated whether the GOP scoring method with the highest correlation to human ratings, GOP\textsubscript{MaxLogit}, effectively identifies mispronounced phonemes. Figure \ref{fig:maxlogit_vs_human} compares phoneme-level mispronunciation error rates predicted by GOP\textsubscript{MaxLogit} with human annotator ratings from the SpeechOcean762 dataset. Discrepancies were further examined by computing the difference in mispronunciation rates between GOP\textsubscript{MaxLogit} predictions and human ratings, highlighting phonemes where the model was overconfident, underconfident, or well-aligned (Figure \ref{fig:maxlogit_vs_human}). 

Phonemes where the GOP\textsubscript{MaxLogit} overestimates mispronunciations (meaning the model assigns significantly higher mispronunciation rates than human raters) include (top 5): /\textipa{@}/, /\textipa{A}/, /\textipa{OI}/, /\textipa{3:}/, and /\textipa{u}/. These phonemes are frequently flagged as mispronounced by the model, despite human raters considering them correctly pronounced in most instances. This suggests that GOP\textsubscript{MaxLogit} is overly confident for these phonemes, possibly misinterpreting slight articulatory variations as errors. Conversely, phonemes where GOP\textsubscript{MaxLogit} \textit{under}estimates mispronunciations (bottom 5), meaning human raters perceive more errors than the model detects, include /\textipa{\ae}/, /\textipa{S}/, /\textipa{eI}/, /\textipa{aU}/, and /\textipa{z}/. Some phonemes exhibit strong agreement between GOP\textsubscript{MaxLogit} predictions and human-rated error rates, indicating well-aligned phoneme classification. These phonemes are /\textipa{b}/, /\textipa{tS}/, /\textipa{i:}/, /\textipa{dZ}/, and /\textipa{A:}/.

\section{Discussion and Conclusion}

In this work, we have analyzed differences in probability-based and logit-based GOP for pronunciation assessment across two datasets, MPC and SpeechOcean762. To answer our RQ, 
our findings indicate that logit-based methods achieve a better classification performance than probability-based GOP; however, their effectiveness depends on the characteristics of the dataset.
GOP\textsubscript{DNN} consistently obtains high recall, but low precision, indicating its tendency to over-detect mispronunciations, as seen in its high overlap between correctly pronounced and mispronounced phonemes. In contrast, logit-based methods (GOP\textsubscript{MaxLogit}) demonstrate better phoneme separation (Fig. \ref{fig:gop_violin}).

A key insight is that GOP\textsubscript{MaxLogit} aligns best with human ratings, achieving the highest PCC scores (Table \ref{tab:mpc-table}), while GOP\textsubscript{DNN}, despite strong classification performance, does not correlate well with expert judgments (Table \ref{tab:speechocean-table}). This suggests that maximum logit values better capture pronunciation quality from a perceptual standpoint. GOP\textsubscript{VarLogit} shows high variability (Table \ref{tab:mpc-table} and \ref{tab:speechocean-table}), making it less reliable, while GOP\textsubscript{Combined} balances probability and logit-based information but still shows some overlap. 

To the best of our knowledge, the highest PCC score reported on the SpeechOcean762 dataset is 0.69 \cite{chao23_interspeech}. However, this was achieved using a multidimensional MDD model that calculates GOP scores while incorporating additional aspects of speech in the SpeechOcean762 dataset. In contrast, our approach focuses solely on logit-based GOP scoring, making direct comparisons with these other methodologies challenging. Future research should focus on reducing reliance on forced alignment, which can introduce errors due to acoustic variability in child and non-native speech, contributing to high recall but low precision. 

In conclusion, our logit-based methodology offers a model-agnostic framework for any CTC-based acoustic model using a threshold-based approach. However, results vary across datasets, with GOP\textsubscript{Margin} performing best on MPC and GOP\textsubscript{MaxLogit} on SpeechOcean762, underscoring the role of acoustic variability.




\section{Acknowledgements}
This publication is part of the project Responsible AI for Voice Diagnostics (RAIVD) with file number NGF.1607.22.013 of the research programme NGF AiNed Fellowship Grants which is financed by the Dutch Research Council (NWO).

\bibliographystyle{IEEEtran}
\bibliography{mybib}

\end{document}